\documentclass[pre,superscriptaddress,twocolumn,showpacs]{revtex4}
\usepackage{graphicx,bbm}
\usepackage{hyperref}
\usepackage{amsmath}
\def\etal#1{ {\em et al.}}
\def\tit#1{}
\def\ii{{\rm i}}

\def\ii{{\rm i}}

\makeatletter
\def\underbracket{\@ifnextchar [ {\@underbracket} {\@underbracket [\@bracketheight]}}
\def\@underbracket[#1]{\@ifnextchar [ {\@under@bracket[#1]} {\@under@bracket[#1][0.4em]}}
\def\@under@bracket[#1][#2]#3{
           \mathop {\vtop {\m@th \ialign {##\crcr $\hfil \displaystyle {#3}\hfil $%
                              \crcr \noalign {\kern 3\p@ \nointerlineskip }\upbracketfill {#1}{#2}
                              \crcr \noalign {\kern 3\p@ }}}}\limits}
\def\upbracketfill#1#2{$\m@th \setbox \z@ \hbox {$\braceld$}
                  \edef\@bracketheight{\the\ht\z@}\bracketend{#1}{#2}
                  \leaders \vrule \@height #1 \@depth \z@ \hfill
                  \leaders \vrule \@height #1 \@depth \z@ \hfill \bracketend{#1}{#2}$}
\def\bracketend#1#2{\vrule height #2 width #1\relax}
\def\downbracketfill#1#2{$\m@th \setbox \z@ \hbox {$\braceld$}
                  \edef\@bracketheight{\the\ht\z@}\downbracketend{#1}{#2}
                  \leaders \vrule \@height #1 \@depth \z@ \hfill
                  \leaders \vrule \@height #1 \@depth \z@ \hfill
\downbracketend{#1}{#2}$}
\def\downbracketend#1#2{\vrule depth #2 width #1\relax}
\makeatother

\begin{document}
\pagenumbering{arabic}

\title{Parametric number covariance in quantum chaotic spectra} 
\author{Vinayak}
\email{vinayaksps2003@gmail.com}
\affiliation{School of Physical Sciences, Jawaharlal Nehru University, New Delhi- 110067}
\author{Sandeep Kumar}
\email{sandeepsps@gmail.com}
\affiliation  {School of Physical Sciences, Jawaharlal Nehru University, New Delhi- 110067}
\affiliation{Department of Physics, H. N. B. Government PG college, Naini, Allahabad- 211008}
\author{Akhilesh Pandey}
\email{ap0700@mail.jnu.ac.in,\\apandey2006@gmail.com}
\affiliation {School of Physical Sciences, Jawaharlal Nehru University, New Delhi- 110067}

\begin{abstract}
We study spectral parametric correlations in quantum chaotic systems and introduce the number covariance as a measure of such correlations. We derive analytic results for the classical random matrix ensembles using the binary correlation method and obtain compact expressions for the covariance. We illustrate the universality of this measure by presenting the spectral analysis of the quantum kicked rotors for the time-reversal invariant and time-reversal non-invariant cases. A local version of the parametric number variance introduced earlier is also investigated.  
\end{abstract}

%

\maketitle
\renewcommand*\thesection{\Roman{section}}
\renewcommand*\thesubsection{\thesection.\Roman{subsection}}

\section{Introduction}
 Random matrix theory (RMT) has been applied in physics as well as in various other scientific disciplines \cite{Mehta, apRMP,BG-Article,Haake, BenRMP, Akemann, Efetov,Guhr:Review}. In physics, the most notable applications of RMT are found in statistical nuclear physics, quantum chaotic systems, and mesoscopic and disordered systems. RMT facilitates a theoretical understanding of the spectral correlations of a physical or a model complex system. An important aspect of RMT is the universality of spectral correlations,  making the results of the classical RMT ensembles, viz. the Gaussian ensembles, useful in many fields. 
  
{Parameter-dependent RMT models \cite{Dyson:1962, ap81} also yield universal results \cite{ ap81,PandeyMehta:1983,fkpt:88, TGW:1990, aprps91, DM, ap1995}. These models are applicable to complex systems in which spectral statistics is governed by an external parameter. In these models one may also consider spectral cross correlations at different parameter values \cite{Efetov,Guhr:Review, BenPRL:93}. Such spectral correlations are associated with the level motion with respect to the parameter, and they are referred to as parametric level correlations.} These correlations have been studied extensively, and their universality {has been tested} in diverse systems, e. g., a hydrogen atom in a uniform magnetic field with the strength of the field as the parameter, resonances in quartz blocks at a uniform temperature where the temperature is an external parameter, and chaotic billiards where Aharonov-Bohm flux or the background potential or boundary parameters are treated as an external parameter \cite{Berry, Simons_etal:93}. In these studies, analytic results for the density-density correlation function \cite{Efetov,Guhr:Review, BenPRL:93} are {of} fundamental {importance}, and they have been obtained using the supersymmetric nonlinear $\sigma$-model for disordered systems. Recent studies of the parametric spectral cross-form factor and the fidelity \cite{Kohler1,Kohler2,Gorin}  reemphasized the importance of parametric correlations.  In addition to the above references we also mention Ref. \cite{ ZaferAlt:93}, which made important contributions to the study of parametric correlations.  

To estimate how long the correlations are sustained in a parameter-driven complex system, it is suggestive to study integrated measures such as a number variance. In this context such a measure appears in the literature \cite{Berry}. However, in our opinion, it is a non-local measure as it involves variance of the staircase function from the ground state. This motivated us to introduce the number covariance as a local measure to study the parametric correlations. It is defined as the covariance of the number of energy levels in intervals of fixed length between spectra for two values of the parameter. By definition it is local, and thus it fulfills the basic requirement for applying the RMT. The number covariance can be calculated numerically from the above-mentioned density-density correlation function, which is known in the form of a multiple integral, or from the spectral cross-form factor which is somewhat simpler. It is surprising that while many measures have been used in this context, the number covariance, which is a natural quantity to use for comparison with numerical data, has not been investigated. 

In this paper we consider the binary correlation method \cite{apRMP,fkpt:88} to derive compact expressions for the number covariance. It turns out to be very close to the results obtained from numerical integrations of the exact correlation functions. We show that our results agree extremely well with the number covariance calculated for the spectra of quantum kicked rotors introduced in \cite{Izrailev}. We also consider a local version of the measure proposed in \cite{Berry}.

\section{Parametric Gaussian Ensembles and The Number Covariance}
We consider the three invariant Gaussian ensembles (GEs) of Hermitian matrices $\mathsf{H}$ of dimension $N$, viz. the Gaussian orthogonal ensemble (GOE), the Gaussian unitary ensemble (GUE), and the Gaussian symplectic ensemble (GSE). We use the Dyson index $\beta$, where $\beta=1,\,2$ and $4$ respectively for these ensembles \cite{apRMP,Mehta}. The joint probability density of matrix elements is given by $P(\mathsf{H}) \propto \exp(-\text{tr} \mathsf{H}^{2}/4v_{\beta}^{2})$. Here the $v_{\beta}^{2}$ are the variances for $\beta$ distinct classes of the off-diagonal matrix elements. Parametric variations in the GEs are described with respect to a parameter $\alpha$ by the ensembles of matrices, $\mathsf{H}_{\alpha}$, defined as $\mathsf{H}_{\alpha} = (\mathsf{H}_{0}+\alpha {V})/\sqrt{1+\alpha^2}$. Here both $\mathsf{H}_{0}$ and ${V}$ belong to the same invariance class of the Gaussian ensembles, and they are independently distributed. It is worth pointing out that similar models are defined for the crossover ensembles \cite{ap81, PandeyMehta:1983,fkpt:88, TGW:1990} with the matrices corresponding to different symmetry classes. Variance $v_{\beta}^{2}$ is the same for $\mathsf{H}_{0},\,\mathsf{H}_{\alpha}$, and ${V}$. Thus $\mathsf{H}_{\alpha}$ and $\mathsf{H}_{0}$ are identically distributed Gaussian ensembles with correlation coefficient $\eta=(1+\alpha^{2})^{-1/2}$ between $H_{0;jk}$ and ${H}_{\alpha;jk}$ for all $j,k$. The scale of the spectral statistics is supplied by $v_{\beta}^{2}$, which we fix by $\beta v_{\beta}^2 N=1$ \cite{apRMP,MonFrench}. In the limit of large $N$, the ensembles-averaged spectral density, $\overline{\rho}(x)$, is described by Wigner's semicircle law \cite{apRMP,ap81}, $\overline{\rho}(x)=\pi^{-1}\sin \psi(x)$, where $\psi(x)=\pi-\cos^{-1}(x/2)$. Notice that the same density is valid for all $\alpha$ and $\beta$.

We introduce the number covariance, $\Sigma^{1,1}_{(\beta)}(x,y;\alpha)$, which is defined as the covariance of the number of levels in the interval $[x,y]$ for $\mathsf{H}_{\alpha}$ and $\mathsf{H}_{0}$. In the limit of large $N$, the number covariance becomes a function of $r$ and $\Lambda$, where $r=|x-y|\overline{\rho}N$ is the average number of eigenvalues in $[x,y]$, and $\Lambda$ is the rescaled parameter defined by \cite{ap81}
\begin{equation}\label{alphaLambda}
\Lambda=\alpha^{2}v_{\beta}^{2}/\overline{D}^{2}=\beta^{-1}\alpha^{2}N\overline{\rho}^{2},
\end{equation}
where $\overline{D}\equiv 1/N\overline{\rho}$ is the average level spacing. We remark that $\Lambda$ depends on $x$ since $\overline{\rho}$ depends on $x$. It has been shown in transition studies \cite{ap81,  PandeyMehta:1983,fkpt:88} that, for $N\to \infty$ and $\alpha\to 0$, the transition in the two-point correlation is abrupt as a function of $\alpha$ but smooth with respect to $\Lambda$. In terms of $r$ and $\Lambda$, the number covariance, $\Sigma^{1,1}_{(\beta)}(r;\Lambda)$, is given by
\begin{equation}
\Sigma^{1,1}_{(\beta)}(r;\Lambda) = \overline{n_{0}(r)n_{\Lambda}(r) } - 
\overline{
n}_{0} (r)
\,
\overline{n}_{\Lambda}(r),
\end{equation}
where {\it overbar} denotes ensemble averaging. $n_{\Lambda}(r)$ is the number of eigenvalues in the interval $[x,y]$ at parameter value $\Lambda$. Notice that the number variance is given by $
\Sigma^{2}_{(\beta)}(r)=\Sigma^{1,1}_{(\beta)}(r;0)$. We also introduce the parametric number variance (PNV), as
\begin{equation}\label{PNV-NCOV}
{V}_{(\beta)}(r;\Lambda)= \overline{(n_{\Lambda}(r) - n_{0}(r))^{2}}=2(\Sigma^{2}_{(\beta)}(r)-\Sigma^{1,1}_{(\beta)}(r;\Lambda)).
\end{equation}
This is the local equivalent of PNV introduced in \cite{Berry}. Note that $\overline{n}=r$ in Eq. (\ref{PNV-NCOV}) whereas $\overline{n}=\mathcal{O}(N)$ in \cite{Berry}. For $\Lambda\to\infty$, ${V}(r;\Lambda)$ becomes $2\Sigma^{2}(r)$, whereas in the latter case, it diverges as $\text{log}(N)$ \cite{Dyson3:1962,fmp:1978}, confirming thereby the nonlocality.


\section{The Binary Correlation Method for the Number Covariance}
To derive the number covariance, we consider the two-point correlation function, $S^{\rho}_{\alpha}(x,y)$, defined as
\begin{equation}\label{srho}
S^{\rho}_{\alpha}(x,y)=\overline{\rho_{\alpha}(x)\rho_{0}(y)}-\overline{\rho}_{\alpha}(x)\,\overline{\rho}_{0}(y),
\end{equation}
where $\rho_{\alpha}(x)$ is the density at parameter value $\alpha$ with the ensemble-averaged density, $\overline{\rho}_{\alpha}(x)$, given by the semicircle law above. Then
\begin{equation}\label{Sig-srho}
\Sigma^{1,1}_{(\beta)}(x,y;\alpha)=\int _{y}^{x}\int _{y}^{x} S^{\rho}_{\alpha}(x',y') dx'dy'.
\end{equation}
The binary correlation method has been described in detail in Refs. \cite{apRMP,fkpt:88,MonFrench}. In this method, the two-point function, $S^{\rho}_{\alpha}(x,y)$, is evaluated in terms of its moments given, for large $N$, as  
\begin{eqnarray}\label{Mom-Srho}
& &
\overline{\langle \mathsf{H}^{p}_{\alpha}\rangle \langle{ \mathsf{H}^{q}_{0}\rangle}}
     - \overline{\langle \mathsf{H}^{p}_{\alpha}\rangle}\,\,   
      \overline{\langle \mathsf{H}^{q}_{0} \rangle}=
    \sum_{\zeta\ge 1}
 \overline{\langle\underbracket[1.0pt]{ \mathsf{H}^{p}_{\alpha}\rangle \langle \mathsf{H}^{q}_{0}}_{\zeta}\rangle}
 \nonumber
 \\
 &\simeq&
 \dfrac{2}{\beta N^{2}} \sum_{\zeta
\ge 
1}
 \zeta\mu_{\zeta}^{p}\mu_{\zeta}^{q} \eta^{\zeta}, ~~~\text{where,}~~\mu_{\zeta}^{p}
=\binom{p}{\dfrac{p-\zeta}{2}}.   
\end{eqnarray}
Here for any $N\times N$ Hermitian matrix $\mathsf{H}$, $\langle\,\mathsf{H}\,\rangle=(\text{tr}\mathsf{H})/N$ denotes the spectral averaging. The first equality in the above equation is exact and denotes a decomposition of summation in terms of $\zeta\, \mathsf{H}_{\alpha}$'s in the first trace, which are cross-correlated with $\zeta\, \mathsf{H}_{0}$'s in the second trace. As in \cite{apRMP,fkpt:88}, the underbracket, together with $\zeta$ underneath, is used to denote these pairs. The second equality is valid for large $N$. $\mu^{p}_{\zeta}$ gives the number of correlated pairs that can be put in the first trace with fixed positions of $\zeta\, \mathsf{H}_{\alpha}$'s; Similarly $\mu^{q}_{\zeta}$ is the number for the second trace.  Also for large $N$, the $\zeta$ cross-correlations appear in a cyclic order. As in \cite{apRMP,fkpt:88,fmp:1978}, $\mu_{\zeta}^{p}$ is the moment of a weighted polynomial:  
\begin{eqnarray}\label{mupzeta}
\mu_{\zeta}^{p}
&=&-\dfrac{1}{\zeta}\int x^{p} \dfrac{d}{dx}\lbrace \overline{\rho}(x)\nu_{\zeta-1}(x) \rbrace dx,\nonumber\\
\nu_{\zeta}(x)&=&(-1)^{\zeta} \dfrac{\sin[(\zeta+1)\psi(x)]}{\sin[\psi(x)]},
\end{eqnarray}
where $\nu_{\zeta}(x)$ is the Chebyshev polynomial of the second kind of order $\zeta$ which is valid for $-2\le x\le2$ with the weight function $\overline{\rho}(x)$. The summation in Eq. (\ref{Mom-Srho}) is valid for $p+q=\text{even}$ and restricted to $\zeta$ such that $p-\zeta=\text{even}$ and $q-\zeta=\text{even}$. Note that $\eta$ carries the entire $\alpha$ dependence of the moments. Finally, carrying out the moment inversion, for large $N$, we find
\begin{equation}
\label{Srho-Assym}
    S^{\rho}_{\alpha}(x,y)\simeq  \frac{\sum_{\zeta\ge 1} \eta^{\zeta}\zeta \cos[\zeta\psi(x)]\cos[\zeta\psi(y)]}{4\beta\pi^{2}N^{2}\sin[\psi(x)]\sin[\psi(y)]}.
\end{equation}

We are interested in local quantities defined in the large-$N$ limit. For instance the unfolded cluster function, $Y^{(\beta)}_{11}(r;\Lambda)$, and the spectral cross-form factor, $\mathcal{K}^{(\beta)}(k;\Lambda)$, are defined by
\begin{eqnarray}\label{Y11}
&&  \frac{ S^{\rho}_{\alpha}(x,y)}{\overline{\rho}(x)\overline{\rho}(y)}
=
-Y^{(\beta)}_{11}(r;\Lambda) 
= \int_{-\infty}^{\infty} \mathcal{K}^{(\beta)}(k;\Lambda)\text{e}^{-2\pi \ii k r} dk.
\nonumber\\
\end{eqnarray}
The cross-form factor has been useful in the semi-classical study \cite{Berry, KuiperMS}, in calculating {the} current correlator \cite{BenPRL:93}, and also in the fidelity analysis \cite{Kohler1,Kohler2}. Note that for $\Lambda=0$, $Y^{(\beta)}_{11}$ and $\mathcal{K}^{(\beta)}$ give respectively the unfolded cluster function and the spectral form factor of the corresponding GEs, viz., $Y^{(\beta)}_{2}(r)-\delta(r)$, and $1-b_{2}^{(\beta)}(k)$,  as defined in \cite{Mehta}. To obtain $Y^{(\beta)}_{11}$ from  Eqs. (\ref{Srho-Assym}, \ref{Y11}), we replace the summation by an integral, using $\zeta=4 N \pi^{2}\overline{\rho}^2 |k|$. Ignoring the rapidly oscillating part of the $\cos[\zeta\Psi(x)]\cos[\zeta\Psi(y)]$ term in Eq. (\ref{Srho-Assym}) we finally get
\begin{eqnarray}\label{Kbeta_Fourier}
 -Y^{(\beta)}_{11}(r;\Lambda)  
&\simeq &
\int_{-\infty}^{\infty} \frac{2|k|}{\beta} 
\text{e}^{-2\beta \pi^{2}\Lambda |k|} \text{e}^{2\pi \ii r k}dk. 
\end{eqnarray}
Note that $2|k|/\beta$ is the small $|k|$ expansion of $\mathcal{K}^{(\beta)}(|k|;0)$. As in \cite{Berry}, to improve the approximation we can replace this term by $\mathcal{K}^{(\beta)}(k;0)$. Comparison of the resulting equation with Eq. (\ref{Y11}) yields,
\begin{equation}\label{Bin_Kbeta}
\mathcal{K}^{(\beta)}(k;\Lambda)\simeq \mathcal{K}^{(\beta)}(k;0) \exp(-2\beta \pi^{2}\Lambda |k|).
\end{equation}
See also \cite{ap1995}, where similar results have been given for the crossover ensembles. In an alternative method used in \cite{apRMP, fkpt:88}, the summation in Eq. (\ref{Srho-Assym}) can be evaluated using an exponential cut-off factor. We introduce $\varepsilon$, replacing $\eta$ by $\eta'$ as $\eta'=\eta\exp[-\varepsilon/2N\pi^{2}\overline{\rho}^{2}]$ in Eq. (\ref{Srho-Assym}), to obtain
\begin{equation}\label{Bin-ep-Kbeta}
\mathcal{K}^{(\beta)}(k;\Lambda)\simeq \frac{2|k|}{\beta}\exp[-2(\beta \pi^{2}\Lambda+\varepsilon)|k|].
\end{equation}

It follows from the stationarity of $S^{\rho}$ that the number covariance, which is a double integral as in (\ref{Sig-srho}), can be written as
\begin{eqnarray}\label{Sig11_K}
\Sigma^{1,1}_{(\beta)}(r;\Lambda)
&=&
 - \int_{-r}^{r} (r-|s|) Y^{(\beta)}_{11}(s;\Lambda)ds\nonumber\\
&=&\int_{-\infty}^{\infty}\mathcal{K}^{(\beta)}(k;\Lambda)
\Bigg[\dfrac{\sin^{2}(\pi k r)}{(\pi k)^{2}}\Bigg]dk.
\end{eqnarray}
Using Eq. (\ref{Bin-ep-Kbeta}) in the second equality of Eq. ({\ref{Sig11_K}}) we get the compact answer,
\begin{equation}\label{Rncov-unf}
\Sigma^{1,1}_{(\beta)}(r;\Lambda)
\simeq
\dfrac{1}{\beta \pi^{2}}\text{ln}\left[
1+\dfrac{r^{2}\pi^{2}}{(\beta \pi^{2} \Lambda+\varepsilon)^{2}}\right].
\end{equation}
The cut-off term has to be fixed with respect to the $\Lambda=0$ result, i.e. $\Sigma^{2}_{(\beta)}(r)$. Since this term has small variation with respect to $r$, we fix its value at which $\Sigma^{1,1}_{(\beta)}(r;0)$ in (\ref{Rncov-unf}) fits the exact $\Sigma^{2}_{(\beta)}(r)$ for large $r$. We find $\varepsilon=0.3676, 0.1035$, and $0.0149$, respectively for $\beta=1,2$ and $4$. For these values of $\varepsilon$ we find that both of our approximations (\ref{Bin_Kbeta}, \ref{Bin-ep-Kbeta}) are close to each other for small $|k|$. Finally we remark that the above result is valid for $r\gtrsim 1$. 

\section{ The parametric number variance}
PNV can be calculated from Eq. (\ref{PNV-NCOV}) along with Eq. (\ref{Sig11_K}) for finite $r$. For $r\to \infty$ and finite $\Lambda$, we find
\begin{equation}\label{BC-PNV}
{V}_{(\beta)}(\infty;\Lambda)
= \int_{-\infty}^{\infty} 
\frac{ \mathcal{K}^{(\beta)}(k;0)-\mathcal{K}^{(\beta)}(k;\Lambda) }
 {(\pi k)^{2}} dk.
\end{equation}
Using Eqs. (\ref{PNV-NCOV},\ref{Rncov-unf}) we obtain,
\begin{equation}\label{BC-PNV1}
{V}_{(\beta)}(\infty;\Lambda)
\simeq
\frac{4}{\beta\pi^{2}}\text{ln} \left( \frac{\varepsilon+\beta\pi^{2}\Lambda}{\varepsilon}\right).
\end{equation}
We remark that the result given in \cite{Berry} is half of our result in Eq. (\ref{BC-PNV}) because their interval $[x,y]$, used in Eq. (\ref{PNV-NCOV}), starts from the ground state. Moreover, they have used the approximation (\ref{Bin_Kbeta}) instead of (\ref{Bin-ep-Kbeta}). 
 

\section{ Exact Results}
The exact results for the density-density correlation function, $\mathsf{R}^{(\beta)}_{11}(r;\Lambda)=1-Y^{(\beta)}_{11}$, and the cross-form factor, $\mathcal{K}^{(\beta)}$, are known \cite{Efetov,Guhr:Review, BenPRL:93}. Note that for $\Lambda=0$, $\mathsf{R}^{(\beta)}_{11}(r;0)$ gives $\mathsf{R}_{2}(r)+\delta(r)$, where $\mathsf{R}_{2}(r)$ is the usual two-level correlation function \cite{Mehta}. These can be used to obtain exact numerical results for $\Sigma^{1,1}_{(\beta)}$. The density-density correlation functions in terms of our above parameter $\Lambda$ are given by
\begin{widetext}
\begin{eqnarray}
\label{para-19}
\mathsf{R}^{(1)}_{11}(r;\Lambda) 
&=&
 1 + \Re \int_{1}^{\infty} dx 
\int_{1}^{\infty}dy \int_{-1}^1 dz  \dfrac{(xy-z)^{2} (1-z^{2})}
{
(x^{2} + y^{2} +z^{2} -2xyz -1)
^{2}} \exp\left[  \ii (\pi r+\ii \delta)(xy-z)\right]\nonumber
\\
&\times & 
\exp\Big[\dfrac{\pi^{2}\Lambda}{2}
(x^{2} + y^{2} +z^{2} -2x^{2}y^{2} -1) 
\Big],
\\
\label{para-20}
\mathsf{R}^{(2)}_{11}(r;\Lambda)
&=&
1 + \int_{0}^{1} dx \int_{1}^{\infty} dy \cos(\pi x r) \cos(\pi y r) 
\exp\left[{\pi^{2}\Lambda(x^{2}-y^{2})}\right], \\
\label{para-21}
{\mathsf R^{(4)}_{11}}(r;\Lambda) 
&=& 1 + \Re \int_{-1}^{1} dx \int_0^{1}dy \int_{1}^{\infty} dz 
 \frac{(xy-z)^{2} (z^{2}-1)}
{
(x^{2} + y^{2} +z^{2} -2xyz -1)
^{2}} 
\exp\left[ - \ii(2\pi r+\ii\delta)(xy-z)\right]
\nonumber
\\
&\times &
\exp\left[-4\pi^{2}\Lambda
(x^{2} + y^{2} +z^{2} -2x^{2}y^{2}-1)
\right],
\end{eqnarray}
\end{widetext}
where $\delta\rightarrow +0$. Using Eqs. (\ref{para-20}) in (\ref{Y11}) one obtains a compact result for $\mathcal{K}^{(2)}(k;\Lambda)$: 
\begin{eqnarray}
\label{para-24}
\mathcal{K}^{(2)}(k;\Lambda)
&=&
\begin{cases}
\exp(-4\pi^{2}\Lambda|k|)\dfrac{\sinh(4\pi^{2}\Lambda k^{2})}{4\pi^{2}\Lambda|k|}~~,& |k|\le 1, \nonumber \\
\exp(-4\pi^{2}\Lambda k^{2})\dfrac{\sinh(4\pi^{2}\Lambda |k|)}{4\pi^{2}\Lambda|k|}~~,& |k|\ge 1.\nonumber\\
\end{cases}
\\
\end{eqnarray}
On the other hand, using (\ref{para-19},\ref{para-21}) in (\ref{Y11}), for $\beta=1$ and $4$,  we get $\mathcal{K}^{(\beta)}(k;\Lambda)$ as double-integrals of the variables $u=xy$ and $v=x^{2}$:
\begin{widetext}
\begin{eqnarray}
\label{para-23}
\mathcal{K}^{(1)}(k;\Lambda)
&=&2k^2\int_{(1,2|k|-1)_>}^{2|k|+1} du\, (1-(u-2|k|)^{2}) \exp(-2\pi^{2}\Lambda u |k|)
\int_{1}^{u^2} dv\,
\dfrac{\exp\left(-\pi^{2}\Lambda(u^{2}-4k^{2}+1-v-u^{2}/v)/2\right)}
{v(u^{2}-4k^{2}+1-v-u^{2}/v)^{2}},\nonumber
\\
\\
\label{para-25}
\mathcal{K}^{(4)}(k;\Lambda)&=&
\dfrac{k^{2}}{4}\int_{(-1,1-|k|)_>}^{1} du\, 
((u+|k|)^{2}-1) \exp(-8\pi^{2}\Lambda u |k|)
\int_{u^{2}}^{1} dv\, \dfrac{\exp(4\pi^{2}\Lambda(u^{2}-k^{2}+1-v-u^{2}/v))}
{v(u^{2}-k^{2}+1-v-u^{2}/v)^{2}}.
\end{eqnarray}
\end{widetext}
One can verify (\ref{Bin_Kbeta}) for small $|k|$, but otherwise the exact results are difficult to deal with analytically. We evaluate $\mathcal{K}^{(1)}(k;\Lambda)$ and $\mathcal{K}^{(4)}(k;\Lambda)$ by solving the double integrals numerically. Next, we use $\mathcal{K}^{(\beta)}(k;\Lambda)$ in Eq. (\ref{Sig11_K}) and evaluate $\Sigma^{1,1}_{(\beta)}(r;\Lambda)$ numerically.

It is worth pointing out that our approximate results Eqs. (\ref{Bin_Kbeta},\ref{Bin-ep-Kbeta}) work well for small $|k|$. However, both approximations yield $\Sigma^{1,1}_{(\beta)}$ close to the exact ones for $r\gtrsim 1$. It comes about because of the $(\pi k)^{-2}$ term which suppresses the contribution of $\mathcal{K}^{(\beta)}(|k|;\Lambda)$ for large $|k|$ in $\Sigma^{1,1}_{(\beta)}(r;\Lambda)$. 

\begin{figure}
        \centering
               \includegraphics [width=0.45\textwidth]{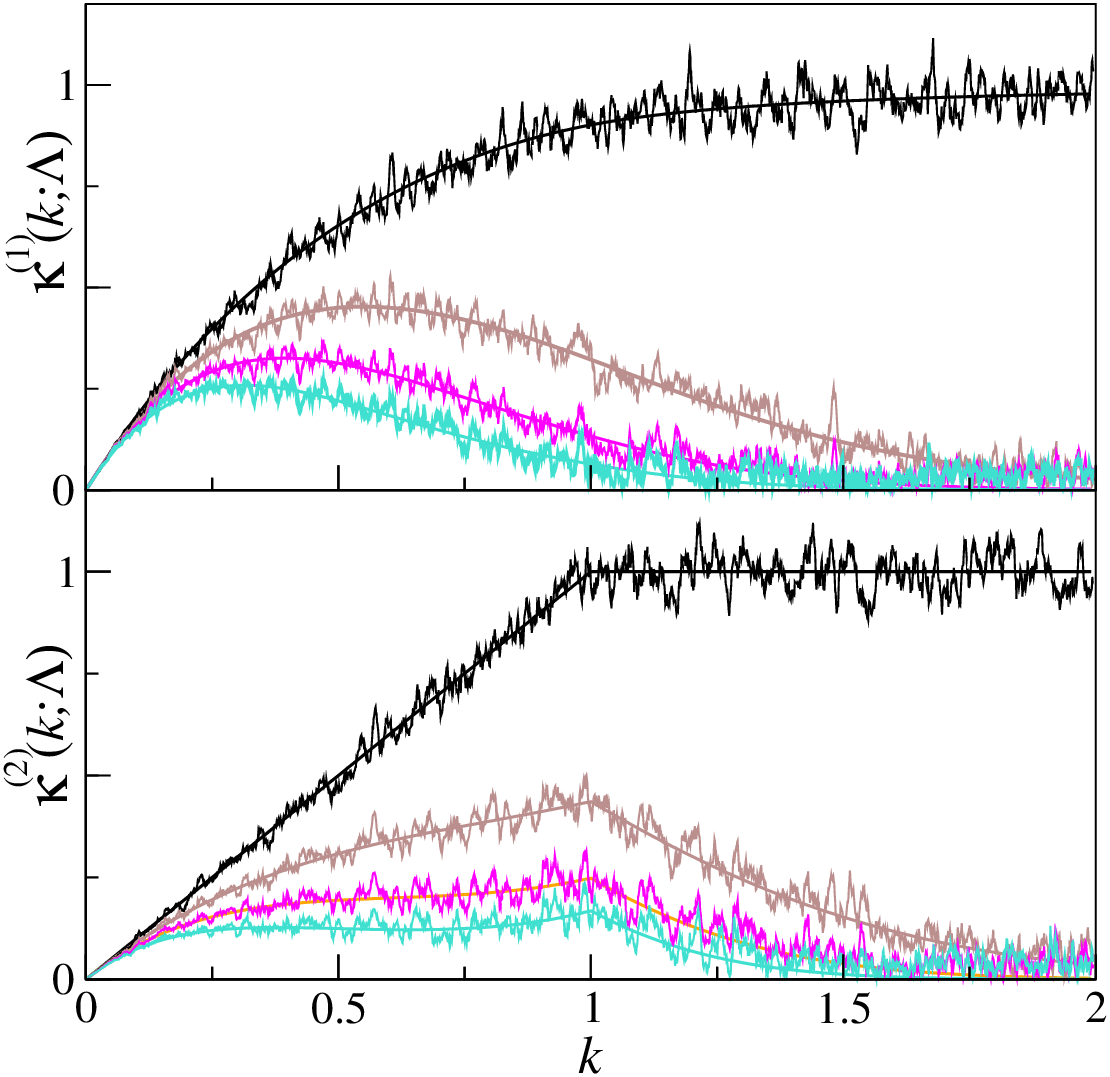}
                \caption{(Color online) The cross-form factor $\mathcal{K}^{(\beta)}(k;\Lambda)$ vs $k$, for $\beta=1$ (top) and $2$ (bottom), at four different values of the parameter $\Lambda$, viz. $\Lambda=0, 0.025, 0.05$ and $0.075$ shown respectively in {\it black} (upper), {\it brown} (mid upper), {\it magenta} (mid lower), and {\it turquoise} (lower). Solid lines represent the exact results and wriggled curves are obtained using Eq. (\ref{Kbeta_Kick}) for eigenangle spectra calculated at $\gamma=0$ (top) and $\gamma=0.1$ (bottom). We have used local averaging in the range $\Delta q=\pm 5$ to reduce the statistical fluctuations. }
\label{Fig-Kbeta}                
\end{figure}

\section{Numerics of the Gaussian Ensembles}
For the GE models, we have considered a $200$-member Gaussian ensemble of $1024$-dimensional $\mathsf{H}_{\alpha}$ matrices for all three $\beta$ at different values of $\alpha$. The variance is fixed such that the semicircle has radius $2$. Since $\Lambda$ depends on $\overline{\rho}(x)$, we choose only $256$ middle levels from each spectrum to ensure that for a given $\alpha$, the density $\overline{\rho}$ and therefore $\Lambda$ do not vary appreciably. 

\begin{figure}
        \centering
               \includegraphics [width=0.45\textwidth]{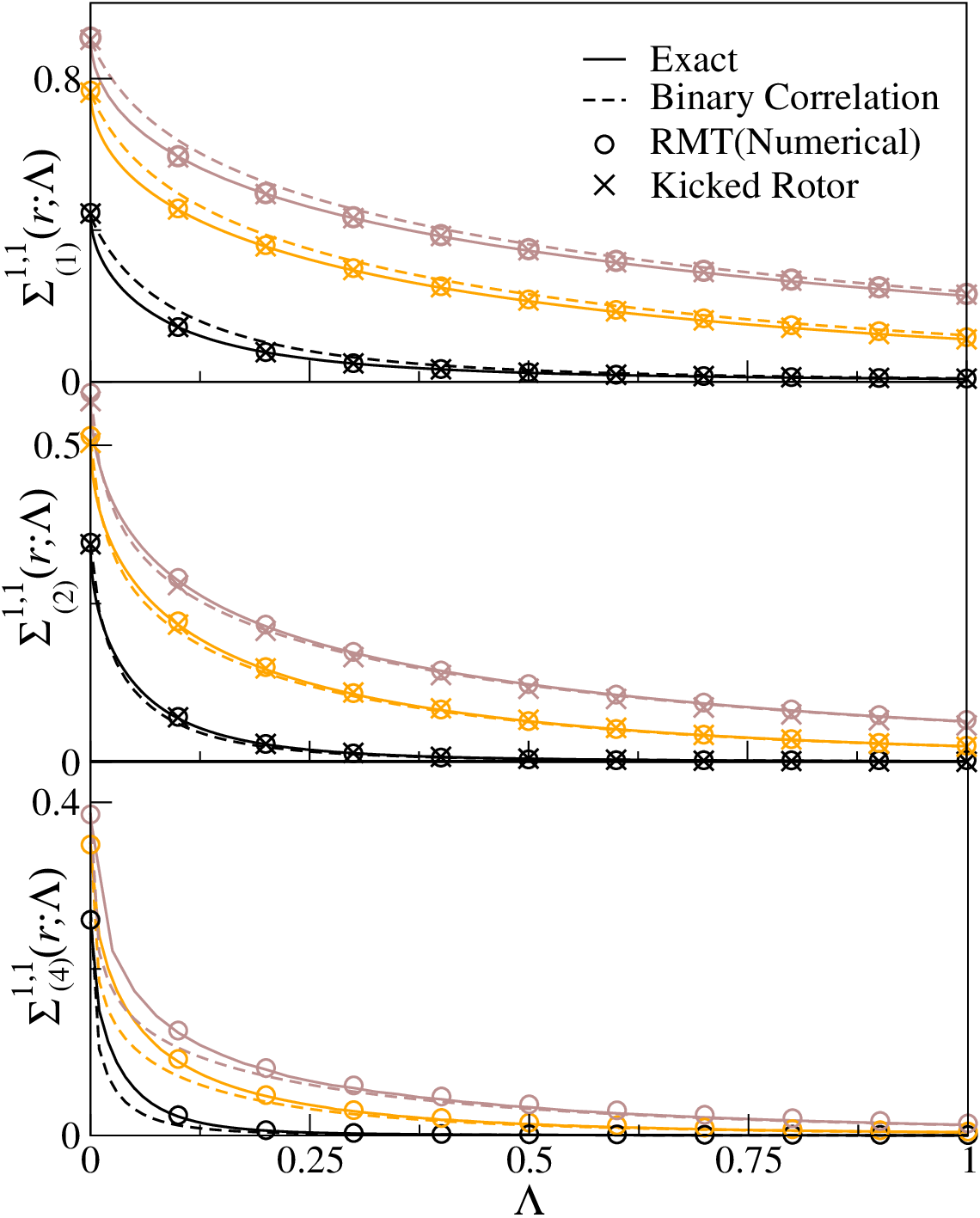}
                \caption{(Color online) The number covariance $\Sigma^{1,1}_{(\beta)}(r;\Lambda)$ vs $\Lambda$, for $\beta=1,2$, and $4$ from top to bottom, at three different values of $r$, viz. $r=1,5$ and $10$ shown respectively in {\it black} (lower), {\it orange} (mid), and {\it brown} (upper). Solid lines represent the exact results obtained from the numerical integration and dashed lines represent approximate results (\ref{Rncov-unf}). Circles represent the RMT data for all three $\beta$ and crosses represent the kicked rotor data for $\beta=1$ and $2$. } 
\label{Fig-ncov}                
\end{figure}

\section{The Quantum Kicked Rotor}
{The quantum Kicked rotor is a prototypical example of quantum chaotic systems.} We consider the eigenangle spectra of quantum kicked rotors \cite{Izrailev,aprps91}. The quantum map is generated in an $N$-dimensional Hilbert space by the time-evolution operator $\mathsf{U}$ of a kicked rotor with torus boundary conditions. The standard case is that of a singly-kicked rotor with $\mathsf{U}=\mathsf{BG}$ where $\mathsf{B}\equiv \mathsf{B}(K)=\exp\left[-\ii K \cos \left(\Theta+\theta_0\right)/\hbar\right]$ and $\mathsf{G}=\exp\left[-\ii\left(\mathsf{p}+\gamma\right)^2/2\hbar\right]$ with $\Theta$ and $\mathsf{p}$ being the position and momentum operators. Here, $K$ is the kicking parameter, $\theta_0$ is the parity-breaking parameter, and $\gamma$ is the time-reversal-breaking parameter $\left(0\le\gamma<1\right)$.We consider parametric correlations arising from small variations $\delta K$ in the kicking strength $K$. Parametric correlation can also be studied with variations in $\theta_{0}$ or $\gamma$ \cite{aprps91}. In the position representation,
$B_{mn}=\exp\left[-\ii\frac{K}{\hbar}\cos\left(\frac{2\pi m}{N}+\theta_0\right)
\right]\delta_{mn}$ and 
$G_{mn}=\frac{1}{N}\sum_{l=-N'}^{N'}\exp\left[-\ii\left(\frac{\hbar}{2}l^2-\gamma l
-\frac{2\pi \mu l}{N}\right) \right]$,
for $\mu=m-n$, $m,n=-N',-N'+1,......,N'$, $N'=\left(N-1\right)/2$ and we set $\hbar=1$. We choose the parameter $\theta_{0}\ne 0$ for parity breaking. For $\gamma=0$, it corresponds to the $\beta=1$ symmetry class, and otherwise it rapidly approaches $\beta=2$. Eigenangle density for the system is constant. The $\Lambda$ parameter is given by \cite{aprps91} $\Lambda={N(\delta K)^2}/{8\beta\pi^2}$, where $\delta K$ is variation in the initial $K$. This can be proved from the first equality of Eq. (\ref{alphaLambda}) by making the correspondence $\alpha\to\delta K$, $\overline{D}\to 2\pi/N$, and $v^{2}\to [\text{tr} \cos^{2}(\Theta+\theta_{0})]/\beta N^{2}=1/2\beta N$. The spectral cross-form factor is calculated as
\begin{equation}\label{Kbeta_Kick}
\mathcal{K}^{(\beta)}(k;\Lambda)=\frac{1}{N}|\overline{\text{tr}\,\mathsf{U}_{\Lambda}^{q}\,\text{tr}\,\mathsf{U}_{0}^{-q}}|,
\end{equation} 
where $q$ is an integer and $k=q/N$.

In numerics we consider $1025$ dimensional matrices $\mathsf{U}$ with $\theta_{0}=\pi/2N$ and $\gamma=0$ and $0.1$ respectively for $\beta=1$ and $2$. Initially $K$ is $10000$ and then varied in small steps of $\delta K\sim 0.1$. This represents one member of the ensemble at different $K$ values. The other independent members of the ensemble are obtained by increasing the initial value of  $K$ in steps of $10000$. Finally, we consider $50$ such members of the ensembles.     

\begin{figure}
        \centering
               \includegraphics [width=0.45\textwidth]{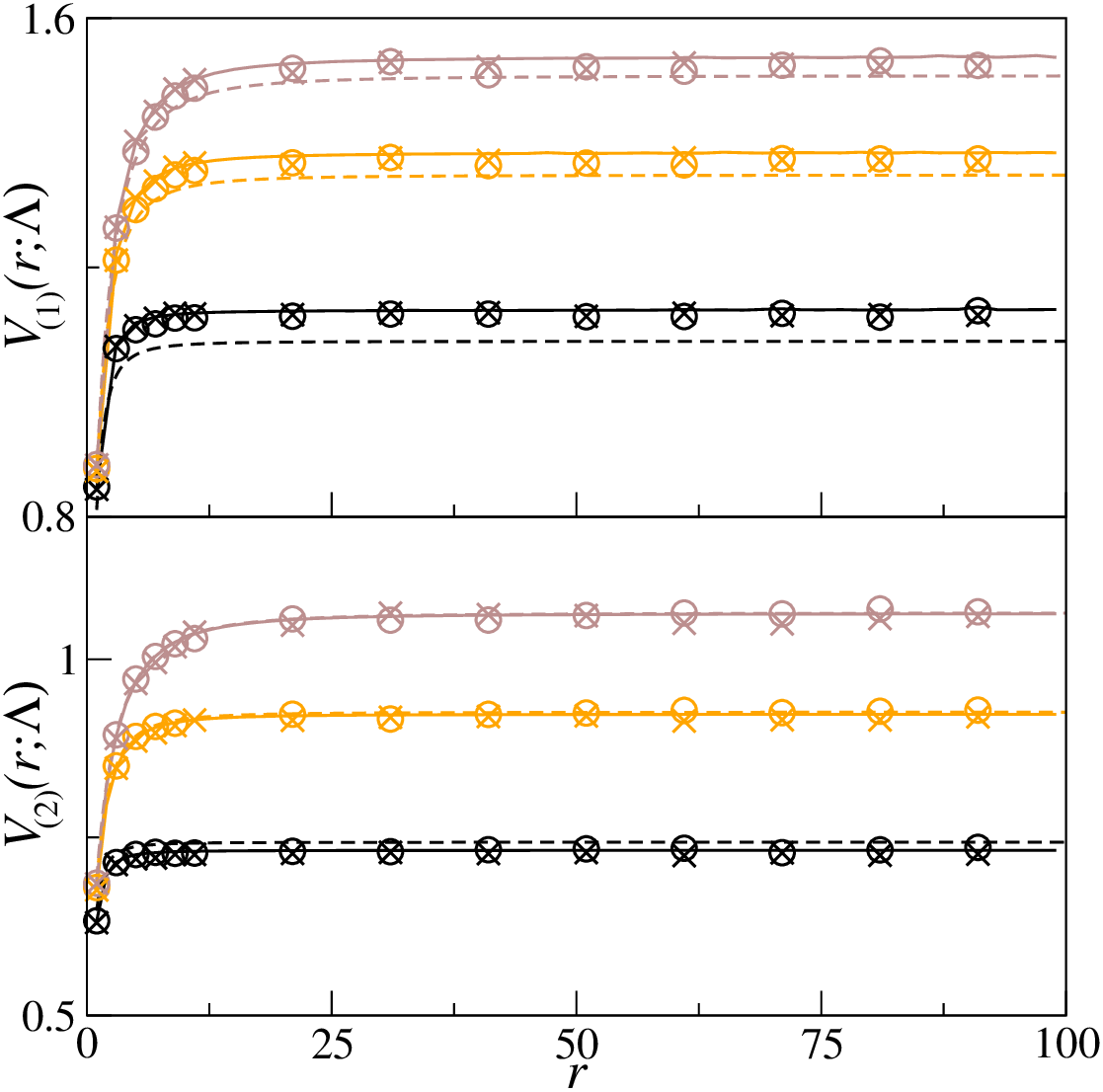}
                \caption{(Color online) PNV ${V}_{(\beta)}(r;\Lambda)$ vs $r$, for $\beta=1$ (top) and $2$ (bottom), at three different $\Lambda$, viz. $\Lambda=0.5, 1$, and $1.5$ for $\beta=1$, and $\Lambda=0.2,0.5$, and $1.0$ for $\beta=2$, shown respectively in {\it black} (lower), {\it orange} (mid), and {\it brown} (upper). As in Fig. \ref{Fig-ncov}, we use lines and symbols for the theory and data respectively.}
\label{Fig-PNV}                
\end{figure}

\section{ Numerical Results}
In Fig. \ref{Fig-Kbeta} we illustrate $\mathcal{K}^{(\beta)}(k;\Lambda)$ vs $k$ for the kicked rotor data evaluated at $\Lambda=0, 0.025,0.05$ and $0.075$. In Fig. \ref{Fig-ncov}, we show $\Sigma^{1,1}_{(1)}(r;\Lambda)$, $\Sigma^{1,1}_{(2)}(r;\Lambda)$, and $\Sigma^{1,1}_{(4)}(r;\Lambda)$ as a function of $\Lambda$ at three values of $r$, viz. $r=1$, $5$, and $10$. Difference in results obtained from the approximations (\ref{Bin_Kbeta}) and (\ref{Bin-ep-Kbeta}) is nominal and therefore the former approximation is not shown. In Fig. \ref{Fig-PNV} we illustrate ${V}_{(1)}(r;\Lambda)$ and ${V}_{(2)}(r;\Lambda)$ as a function of $r$ at several values of $\Lambda$. In this figure we consider $r$ upto $100$. For $r=100$, ${V}_{(\beta)}(r;\Lambda)$ becomes almost independent of $r$.  

It is evident from these figures that exact results are in excellent agreement with the kicked rotor data. Also, our binary correlation results yield a very good approximation to the exact results.

\section{Conclusion}
In conclusion, we have defined the parametric number covariance to study parametric correlations in quantum chaotic spectra. We have shown that the local spectral fluctuations become rapidly independent as the parameter $\alpha$ of the system is varied. Smooth statistical variations are found as a function of a rescaled parameter $\Lambda=\alpha^{2}\overline{\rho}^{2}N/\beta$. For spectra with $\overline{\rho}=\mathcal{O}(1)$, we find $\Lambda=\mathcal{O}(1)$ when $\alpha=\mathcal{O}(N^{-1/2})$. For such small values of $\alpha$ the global correlations between the spectra are close to 1. 

We have dealt with the three $\beta$ cases and derived the number covariance for the Gaussian ensembles, using the binary correlation method, which is close to the results obtained form numerical integration of the exact formula. We have shown its universality in the quantum kicked rotor spectra for time-reversal invariant and time-reversal non-invariant systems.


\end{document}